\documentclass[aps,prb,superscriptaddress,amsmath,twocolumn]{revtex4}

\pdfoutput=1
\usepackage{times}
\usepackage{graphicx}
\usepackage{epsfig}

\begin{document}

\title{Size-consistent variational approaches to non-local pseudopotentials:
standard and lattice regularized diffusion Monte Carlo methods revisited}

\author{Michele Casula}
\affiliation{Centre de Physique Th\'eorique, CNRS,  \'Ecole Polytechnique, 91128 Palaiseau Cedex, France}
\author{Saverio Moroni}
\affiliation{International School for Advanced Studies (SISSA), Via Beirut 2--4, 34014 Trieste, Italy}
\affiliation{INFM Democritos National Simulation Center, Via Beirut 2--4, 34014 Trieste, Italy}
\author{Sandro Sorella}
\affiliation{International School for Advanced Studies (SISSA), Via Beirut 2--4, 34014 Trieste, Italy}
\affiliation{INFM Democritos National Simulation Center, Via Beirut 2--4, 34014 Trieste, Italy}
\author{Claudia Filippi}
\affiliation{Faculty of Science and Technology and MESA+ Research Institute, University of Twente, P.O. Box 217, 7500 AE Enschede, The Netherlands}

\begin{abstract}
We propose improved versions of the standard diffusion Monte Carlo 
(DMC) and the lattice regularized diffusion Monte Carlo (LRDMC) algorithms.
For the DMC method, we refine a scheme recently devised to treat non-local 
pseudopotential in a variational way. We show that such scheme --when
applied to large enough systems-- maintains its effectivness only at 
correspondingly small enough time-steps, and we present two simple upgrades
of the method which guarantee the variational property in a size-consistent
manner.
For the LRDMC method, which is size-consistent and variational by construction,
we enhance the computational efficiency by introducing (i) an improved 
definition of the effective lattice Hamiltonian which
remains size-consistent and entails a small
lattice-space error with a known leading term, and (ii) a new randomization 
method for the positions of the lattice knots which requires a single 
lattice-space.
\end{abstract}

\maketitle

\section{Introduction}
\label{Sec:intro}

The fixed-node diffusion Monte Carlo (DMC) method is often the method of choice for 
accurate computations of many-body systems~\cite{FMNR01}.
Since the scaling of DMC with the number of electrons $N$ is a modest $N^4$, the method
has been employed in recent years to accurately compute electronic properties of large 
molecular and solid systems where conventional highly-correlated quantum chemistry approaches 
are very difficult to apply.  Unfortunately, for full-core atoms the computational cost of DMC increases
approximately~\cite{C86,HRL87} as $Z^{5.5-6.5}$ with the atomic number $Z$. Therefore, the 
use of pseudopotentials is an essential ingredient in the application of DMC to complex
systems to reduce the effective value of $Z$ and significantly improve the efficiency 
of the method.

The use of pseudopotentials in DMC poses however a problem since pseudopotentials are usually
non-local and the non-locality introduces a fermionic sign problem additional to the 
one due to the anti-symmetry of the electronic wave function. The commonly adopted
solution is to ``localize'' the non-local pseudopotential on the trial wave function and use this
local potential in the DMC simulation~\cite{fahy,mitas}. Unfortunately, 
the so-called locality approximation (LA) does not 
ensure variationality and alternative schemes employing a different effective Hamiltonian were recently 
introduced to overcome this difficulty~\cite{CFS05,C06}. 

In the lattice regularized diffusion Monte Carlo (LRDMC) algorithm~\cite{CFS05}, both the Laplacian and the 
non-local pseudopotentials are discretized such that the corresponding imaginary time 
propagator $\langle \exp[-\tau {\cal H}] \rangle$ assumes non-zero
values only on a finite set of points, and the lattice Green function Monte Carlo algorithm can be employed,
that ensures variationality and stability all along the simulation~\cite{ceperley}. 
Alternatively, another scheme which
is based on the standard DMC algorithm was developed~\cite{C06}. The latter exploits the discretization
of the propagator only in the part depending on the non-local pseudopotentials, 
and a non-local effective Hamiltonian is defined in order to
fulfill the fixed node constraint.
Here, we show that this variational DMC scheme 
is however not size consistent at finite time-steps. Indeed, the time-step error 
strongly depends on the system size and, upon increasing the number of particles at fixed time-step, 
the corresponding energies approach those given by DMC with LA.
In this paper, we explain how to cure this problem and 
present a simple formulation of the algorithm
which is size-consistent and suffers at the same time from a smaller time-step error.
Moreover, we define a better discretization rule for the LRDMC effective Hamiltonian, 
which reduces the lattice-space bias, remains size consistent as in the 
original formulation,  
and improves the efficiency of the method.

In Section~\ref{Sec:methods}, we briefly summarize the problems introduced by the use of 
non-local pseudopotentials in the standard DMC and describe in detail the variational DMC algorithm 
of Ref.~\onlinecite{C06} to treat pseudopotentials beyond the commonly used LA.
In Section~\ref{Size_consis}, we present our size-consistent variational approach to 
non-local pseudopotentials in DMC and demonstrate its effectiveness on a series 
of oxygen systems of increasing size. 
In Section~\ref{Sec:lrdmc}, we briefly describe the LRDMC method, which is variational
by its own nature, and give a better prescription for the lattice regularization of the continuous Hamiltonian to
always guarantee a well defined and faster zero lattice-space extrapolation.
Finally, in Section~\ref{Sec:performance}, we discuss the behavior of the discretization error 
of the different DMC algorithms (in time or space as appropriate)
and comment on the relative efficiency of the methods presented here.

\section{DMC and non-local pseudopotentials}
\label{Sec:methods}

In DMC, the projection to the ground state wave function of an Hamiltonian $\cal H$ is 
performed by stochastically applying the operator $\exp[-\tau{\cal H}]$ to a trial wave 
function $\Psi_{\rm T}$. If the projection is formulated in real space and importance 
sampling introduced, the mixed distribution 
$f({\bf R},t)=\Psi_{\rm T}({\bf R})\Psi({\bf R},t)$ is then propagated as
\begin{eqnarray}
f({\bf R}',t+\tau)=\int {\rm d} {\bf R}\, G({\bf R}',{\bf R},\tau)f({\bf R},t)\,,
\end{eqnarray}
where the importance sampling Green's function is defined as
\begin{eqnarray}
G({\bf R}',{\bf R},\tau)&=&\frac{\Psi_{\rm T}({\bf R}')}{\Psi_{\rm T}({\bf R})}\langle {\bf R}'|\exp[-\tau{\cal H}]|{\bf R}\rangle\,.
\end{eqnarray}
The fixed-node (FN) approximation is usually employed for fermionic systems to avoid the collapse 
to the bosonic ground state.
In continuous systems, it is implemented by constraining the diffusion process
within the nodal pockets of the 
trial wave function.  For long times, the distribution $f({\bf R},t)$ approaches
$\Psi_{\rm T}({\bf R})\Psi_{\rm FN}({\bf R})$ where $\Psi_{\rm FN}({\bf R})$ is the ground state wave function 
consistent with the boundary condition that it vanishes at the nodes of $\Psi_{\rm T}$. The FN energy is an
upper bound to the true fermionic ground state energy.

When a non-local potential ${\cal V}^{\rm NL}$ is employed to remove the core electrons, the off-diagonal elements 
of the Hamiltonian in real space are generally non-zero and the standard DMC approach cannot be applied. 
If we analyze the behavior of the propagator at short time-steps:
\begin{eqnarray}
\langle {\bf R}' | \exp[-\tau{\cal H}] | {\bf R} \rangle 
\approx \delta_{{\bf R}',{\bf R}}-\tau\langle {\bf R}' | {\cal H}  | {\bf R} \rangle\,,
\end{eqnarray}
we note that, while the diagonal elements can always be made positive by choosing $\tau$ small enough, the off-diagonal 
elements are positive if and only if the off-diagonal elements of the Hamiltonian are non-positive. This condition is not 
always met in the presence of non-local pseudopotentials, so the fermionic sign problem reappears even if one works
in the FN approximation.
Consequently, in addition to the FN approximation, the LA is commonly introduced where 
the non-local potential ${\cal V}^{\rm NL}$ is replaced by a local quantity ${\cal V}^{\rm LA}$ obtained by ``localizing'' the potential
on the trial wave function:
\begin{eqnarray}
{\cal V}^{\rm LA}({\bf R})=\frac{\langle {\bf R} | {\cal V}^{\rm NL} | \Psi_{\rm T} \rangle}
{\langle {\bf R} |\Psi_{\rm T} \rangle}\,.
\end{eqnarray}
The DMC algorithm in the LA yields the FN ground state of the effective Hamiltonian
${\cal H}^{\rm LA}$ with the local potential ${\cal V}^{\rm LA}$ instead of the original non-local ${\cal V}^{\rm NL}$. 
The fixed-node energy in the LA is equal to
\begin{eqnarray}
{\rm E}^{\rm LA}_{\rm FN}=\frac{\langle \Psi^{\rm LA}_{\rm FN} | {\cal H}^{\rm LA} | \Psi^{\rm LA}_{\rm FN} \rangle}{\langle \Psi^{\rm LA}_{\rm FN} | \Psi^{\rm LA}_{\rm FN} \rangle}\,,
\end{eqnarray}
and estimated by sampling the mixed distribution $\Psi_{\rm T}\Psi^{\rm LA}_{\rm FN}$ as 
\begin{eqnarray}
{\rm E}^{\rm LA}_{\rm FN}
=\frac{\langle \Psi^{\rm LA}_{\rm FN} | {\cal H}^{\rm LA} | \Psi_{\rm T} \rangle}{\langle \Psi^{\rm LA}_{\rm FN} | \Psi_{\rm T} \rangle}
=\frac{\langle \Psi^{\rm LA}_{\rm FN} | {\cal H} | \Psi_{\rm T} \rangle}{\langle \Psi^{\rm LA}_{\rm FN} | \Psi_{\rm T} \rangle}\,.
\end{eqnarray}
Since $\Psi^{\rm LA}_{\rm FN}$ is the fixed-node ground state of ${\cal H}^{\rm LA}$ and not of the original 
Hamiltonian ${\cal H}$, the mixed average energy of ${\cal H}$ is not equal to its expectation value on
the wave function $\Psi^{\rm LA}_{\rm FN}$,
\begin{eqnarray}
\frac{\langle \Psi^{\rm LA}_{\rm FN} | {\cal H} | \Psi^{\rm LA}_{\rm FN} \rangle}{\langle \Psi^{\rm LA}_{\rm FN} | \Psi^{\rm LA}_{\rm FN} \rangle}\,.
\end{eqnarray}
Therefore, $E^{\rm LA}_{\rm FN}$ is in general not an upper bound to the ground state of ${\cal H}$ 
and the variational principle may not apply.

\subsection{Beyond the locality approximation}
\label{Beyond_LA}

The lattice regularized diffusion Monte Carlo (LRDMC) algorithm was recently developed to overcome this 
difficulty~\cite{CFS05} and then extended to the continuum formulation of DMC~\cite{C06}. 
Both algorithms provide a variational scheme to treat non-local pseudopotentials in DMC by introducing an
effective Hamiltonian, different from the one used in the LA approximation, which provides an upper bound
to the ground state of the original Hamiltonian. We briefly describe here the algorithm in the framework
of continuum DMC.

We first apply a Trotter expansion for small time-steps to the importance sampling Green's function,
\begin{eqnarray}
G({\bf R}',{\bf R},\tau)
\approx\int {\rm d}{\bf R}''T^{\rm NL}({\bf R}',{\bf R}'',\tau) G^{\rm loc}({\bf R}'',{\bf R},\tau)\,,
\label{int_form}
\end{eqnarray}
where we have split the Hamiltonian into a local and a non-local operator. The propagator
$G^{\rm loc}({\bf R}',{\bf R},\tau)$ is equal to the drift-diffusion-branching Green's function for the local 
component of the Hamiltonian, 
\begin{eqnarray}
\frac{1}{(2\pi\tau)^{3N/2}}e^{-\left[{\bf R}'-{\bf R}-\tau {\bf V}({\bf R})\right]^2/2\tau}
e^{-\tau E_{\rm L}^{\rm loc}({\bf R}')}\,,
\label{G^loc}
\end{eqnarray}
where the velocity is defined as ${\bf V}({\bf R})=\nabla\Psi_{\rm T}({\bf R})/\Psi_{\rm T}({\bf R})$ and 
$E_{\rm L}^{\rm loc}({\bf R})={\cal H}^{\rm loc}\Psi_{\rm T}({\bf R})/\Psi_{\rm T}({\bf R})$ is the local energy of the 
local part of the Hamiltonian (kinetic ${\cal K}$ plus local potential ${\cal V}^{\rm loc}$). 
The transition $T^{\rm NL}$ contains the non-local potential,
\begin{eqnarray}
T^{\rm NL}({\bf R}',{\bf R},\tau)&=&\frac{\Psi_{\rm T}({\bf R}')}{\Psi_{\rm T}({\bf R})}\langle {\bf R}'|\exp[-\tau{\cal V}^{\rm NL}]|{\bf R}\rangle\nonumber\\
&\approx& \delta_{{\bf R}',{\bf R}}-\tau V_{{\bf R}',{\bf R}}\,.
\label{T^NL}
\end{eqnarray}
where $V_{{\bf R}',{\bf R}}=\Psi_{\rm T}({\bf R}')/\Psi_{\rm T}({\bf R}) \langle {\bf R}'|{\cal V}^{\rm NL}|{\bf R}\rangle$. 
In both the variational Monte Carlo and the standard DMC method with the LA approximation, one adopts a quadrature 
rule with a discrete mesh of points, belonging to a regular polyhedron used 
to evaluate the projection of the non-local component on a given trial wave function. 

Consequently, the number of elements $V_{{\bf R}',{\bf R}}$ is finite
and the transition $T^{\rm NL}$ corresponds to the move of one electron on the grid obtained by considering the union of
the quadrature points generated for each electron and pseudoatom
(center of a non-local pseudopotential). 
Moreover, in order to work with a small quadrature mesh, the 
vertices of the polyhedron are defined in a frame rotated by $\theta$ and $\phi$, 
the azimuthal and planar angle respectively, which are taken randomly 
for each electron.

As discussed above, performing a transition based on $T^{\rm NL}$ poses however a problem since $T^{\rm NL}$ can be negative given that both
$\Psi_{\rm T}({\bf R}')/\Psi_{\rm T}({\bf R})$ and $\langle {\bf R}'|{\cal V}^{\rm NL}|{\bf R}\rangle$ can change sign. 
A solution is to apply the FN approximation not only to $G^{\rm loc}$ but also to $T^{\rm NL}$ by keeping only the transition
elements which are positive:
\begin{eqnarray}
T^{\rm NL}_{\rm FN}({\bf R}',{\bf R},\tau)&=& \delta_{{\bf R}',{\bf R}}-\tau V^-_{{\bf R}',{\bf R}}\,.
\label{T^NL_FN}
\end{eqnarray}
where $V^\pm_{{\bf R}',{\bf R}}=[V_{{\bf R}',{\bf R}}\pm|V_{{\bf R}',{\bf R}}|]/2$. The discarded elements
are included in the so-called sign-flip term, which is then added to the diagonal local potential as
\begin{eqnarray}
{\cal V}_{\rm eff}^{\rm loc}({\bf R})={\cal V}^{\rm loc}({\bf R})+\sum_{{\bf R}'}V^+_{{\bf R}',{\bf R}}\,.
\label{V_eff^loc}
\end{eqnarray}
The resulting effective Hamiltonian ${\cal H}_{\rm eff}$ is therefore given by
\begin{eqnarray}
\langle{\bf R}|{\cal H}_{\rm eff}|{\bf R}\rangle&=& \langle{\bf R}|{\cal K}|{\bf R}\rangle+
{\cal V}_{\rm eff}^{\rm loc}({\bf R})\nonumber\\
\langle{\bf R}'|{\cal H}_{\rm eff}|{\bf R}\rangle&=& \langle{\bf R}'|{\cal V}^{\rm NL}|{\bf R}\rangle
\hspace*{1cm}{\rm if}\ V_{{\bf R}',{\bf R}}<0\,,
\label{H_eff}
\end{eqnarray}
and yields the same local energy as the original Hamiltonian ${\cal H}$. In contrast to
the LA Hamiltonian, its ground state energy is an upper bound to the ground state energy of the true
Hamiltonian. Therefore, the variational principle is recovered and, in addition, the use of ${\cal H}_{\rm eff}$
in combination with the $T_{\rm FN}^{\rm NL}$ transition cures the instabilities which are commonly observed in a DMC run with the
LA Hamiltonian and are due to the negative divergences of the localized potential on the nodes of $\Psi_{\rm T}$.

In the branching term of $G^{\rm loc}$ (Eq.~\ref{G^loc}), the local potential 
${\cal V}^{\rm loc}$ is replaced by ${\cal V}_{\rm eff}^{\rm loc}$ (Eq.~\ref{V_eff^loc}) 
and the weights of the walkers are multiplied by an additional factor which enters in the normalization 
of the transition $T_{\rm FN}$ and to order $\tau$ is equal to:
\begin{eqnarray}
\sum_{{\bf R}'}T_{\rm FN}^{\rm NL}({\bf R}',{\bf R}) \approx\exp\left(-\tau\sum_{{\bf R}'}V^-_{{\bf R}',{\bf R}}\right)\,.
\end{eqnarray}
The weights are therefore given by
\begin{eqnarray}
w=w_{\rm eff}^{\rm loc}\sum_{{\bf R}'}T_{\rm FN}^{\rm NL}({\bf R}',{\bf R}) =\exp\left[-\tau E_{\rm L}({\bf R})\right]\,,
\end{eqnarray}
where $E_{\rm L}({\bf R})={\cal H}_{\rm eff}\Psi_{\rm T}({\rm R})/\Psi_{\rm T}({\rm R})={\cal H}\Psi_{\rm T}({\rm R})/\Psi_{\rm T}({\rm R})$.

The basic algorithm as proposed in Ref.~\cite{C06} therefore consists of the following steps: 
\begin{itemize}
\itemsep0ex
\item[1.] The walker drifts and diffuses 
from ${\bf R}$ to ${\bf R}'$. The move is followed by an accept/reject step as in standard DMC.
\item[2.] The weight of the walker is multiplied by the branching factor 
$\exp\left[-\tau\left(E_{\rm L}({\bf R}')-E_{\rm T}\right)\right]$ where the trial energy $E_{\rm T}$ has been 
introduced.
\item[3.] The walker moves to ${\bf R}''$ according to the transition 
probability $T^{\rm NL}_{\rm FN}({\bf R}'',{\bf R}')/\sum_{{\bf R}'''}T_{\rm FN}^{\rm NL}({\bf R}''',{\bf R}')$.
\end{itemize}
For large systems, the first step is implemented not by moving all the electron together but by sequentially drifting and diffusing
each electron and applying the accept/reject step after each single-electron (SE) move.

\section{Size-consistency}
\label{Size_consis}

In the move governed by the transition $T^{\rm NL}_{\rm FN}$, only one electron is displaced on the grid of the 
quadrature points generated by considering all the pseudoatoms and all the electrons.
Therefore, for given time-step, the probability of a successful move will increase with the system size (i.e.\ the number of
electrons) and saturate to one for sufficiently 
large systems. In this limit, the effect of the move will become independent of the system size and lead to one 
electron being displaced at each step. Therefore, for sufficiently large systems, the overall
impact of the non-local move will decrease and the algorithm will effectively behaves more and more like in the LA procedure.

\begin{figure}[t]
\centering
\includegraphics[width=1.0\columnwidth]{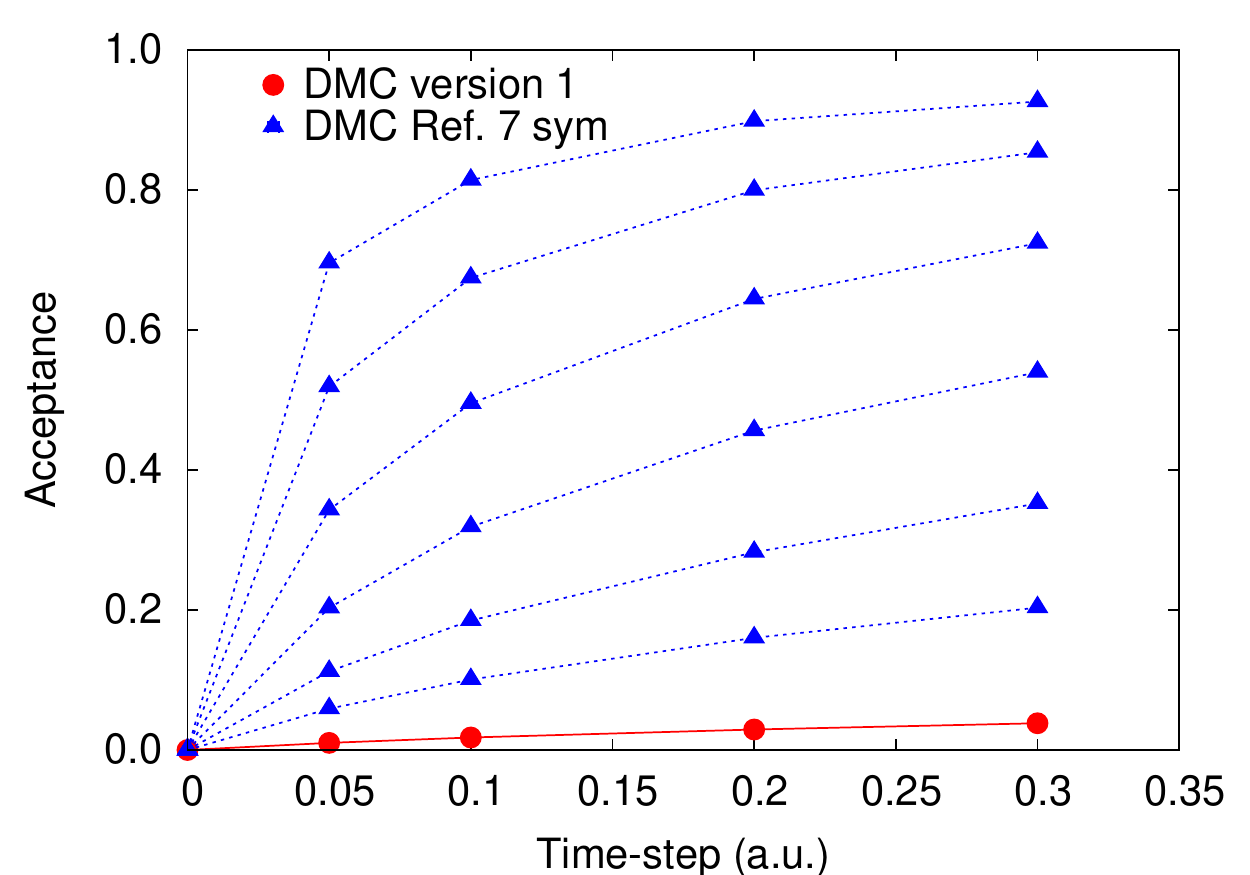}
\caption{Acceptance of the $T_{\rm FN}^{\rm NL}$ move as a function of time-step. The increasing dotted curves 
correspond to systems with 1, 2, 4, 8, 16 and 32 oxygen atoms aligned at a distance of 30~$\AA$ from each other.
The dotted curves are obtained with the algorithm of Ref.~\onlinecite{C06} while the lowest continuum curve is obtained 
with the size-consistent DMC algorithm (version 1) we propose. We only show the size-consistent curve with 1 atom as it is 
indistinguishable from the ones obtained for the larger systems.}
\label{fig:jwalkalize}
\end{figure}

To demonstrate the size-consistency problem of the algorithm as originally formulated in Ref.~\cite{C06}, we consider a series of systems 
consisting of an increasing number $M$ of oxygen atoms aligned 30~$\AA$ apart. 
The oxygen atom is described by an $s$-non-local energy-consistent Hartree-Fock pseudopotential~\cite{BFD07}.
The trial wave function is of the Jastrow-Slater type with a single determinant expressed on a cc-pVDZ 
basis~\cite{BFD07} and a Jastrow factor which includes electron-electron and 
electron-nucleus terms~\cite{FU96}. All Jastrow and orbital parameters are optimized in energy 
minimization~\cite{UTFSH07} for a single atom and the wave function of a system with more than one oxygen is
obtained by replicating the wave function of one atom on the other centers. In Fig.~\ref{fig:jwalkalize}, 
we plot the acceptance of the $T_{\rm FN}^{\rm NL}$ move as a function of time-step for systems containing
1, 2, 4, 8, 16, and 32 oxygen atoms.  For each system size, the probability goes to zero at small time-steps 
and increases for larger values of $\tau$ as expected from the expression of $T_{\rm FN}^{\rm NL}$ (Eq.~\ref{T^NL_FN}).
The acceptance increases with the size of the system; as a function of the time-step, it approaches its asymptotic value of one more quickly for the larger systems.

To better understand the overall behavior of the algorithm with increasing system size, we also analyze
the FN energy as a function of time-step.
We are interested in comparing the results obtained with the conventional LA approach and with the algorithm of 
Ref.~\cite{C06} described in the previous section. For a more meaningful and clear comparison with conventional 
DMC with the LA which employs a symmetrized branching factor, we modify the original algorithm as described in 
the previous section to also use a symmetrized branching factor:
\begin{eqnarray}
\exp\left[-\tau\left(E_{\rm L}({\bf R})+E_{\rm L}({\bf R}')\right)/2\right]
\label{sym_branching}
\end{eqnarray}
where ${\bf R}$ and ${\bf R}'$ are the coordinates before the drift-diffusion move of the first electron and after
the drift-diffusion move of the last electron, respectively, if the electrons are displaced subsequently.
Such a simple modification is allowed as it only entails a different time-step error, which 
we actually find to be significantly smaller than the one obtained with the algorithm of Ref.~\cite{C06} as 
we detail in the Section~\ref{Sec:performance}.

\begin{figure}[t]
\centering
\includegraphics[width=1.0\columnwidth]{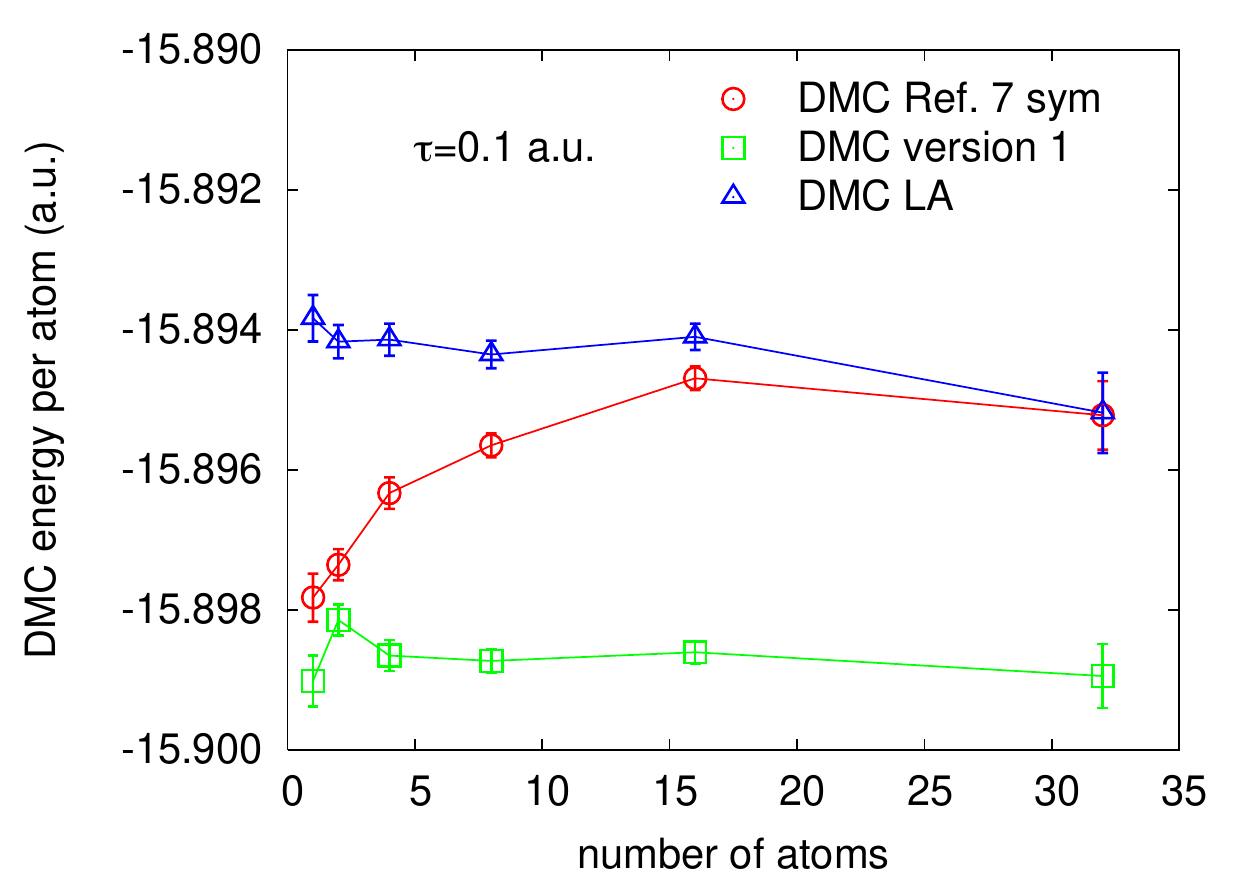}\\
\includegraphics[width=1.0\columnwidth]{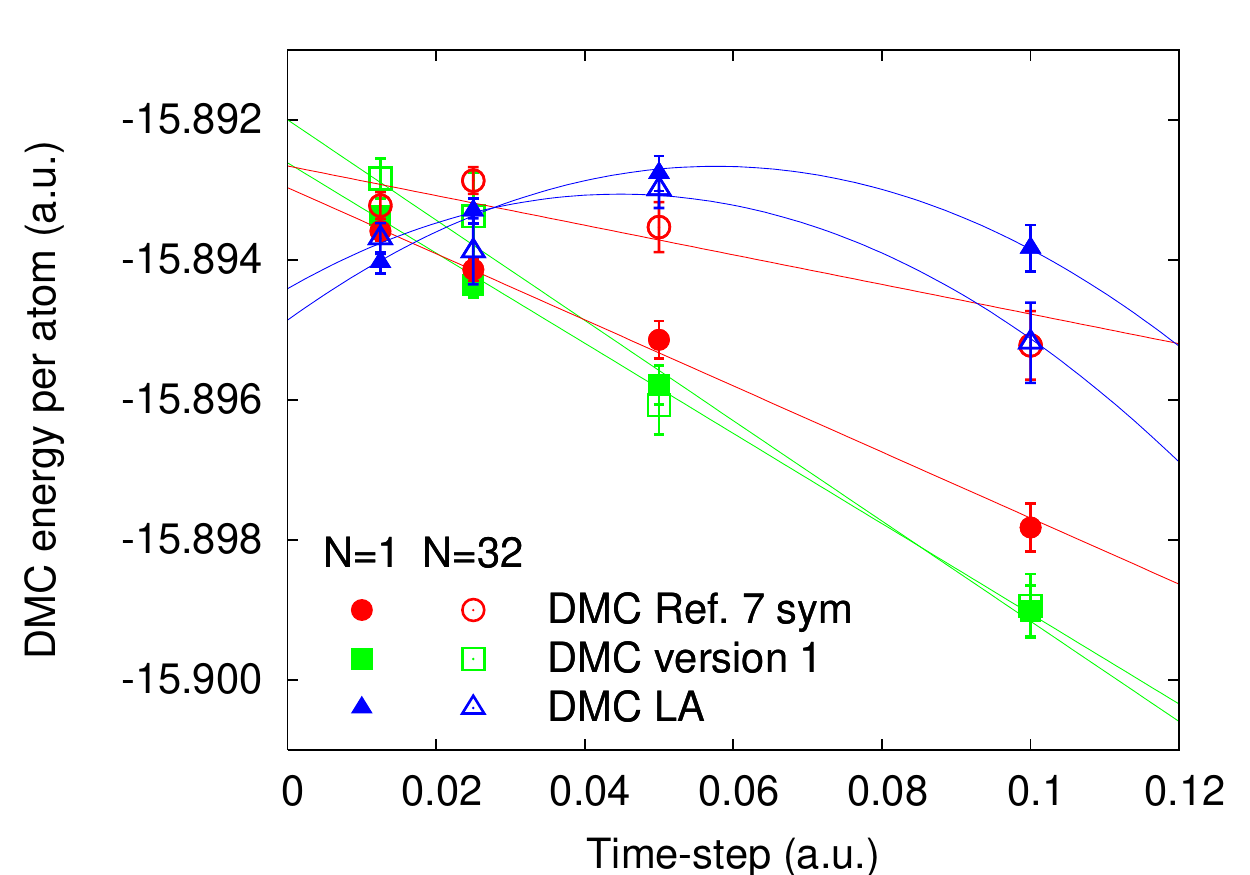}
\caption[]{Upper panel: DMC FN energy per atom for systems of $M$ isolated 
oxygen atoms for $\tau=0.1$. For given time-step, the results of the algorithm 
of Ref.~\cite{C06} (red circles) approach those of the LA (blue triangles) 
upon increasing the system size, whereas the present algorithm (DMC version 1, 
green squares) gives values independent of $M$. In the three algorithms, the 
branching factor is updated after having moved all the electrons (AE 
branching).
Lower panel: time-step dependence of the DMC FN energy for the same three 
algorithms for $M=1$ (filled symbols) and $M=32$ (open symbols). The algorithm 
of Ref.~\cite{C06} has a problematic extrapolation to zero time-step for large 
enough system size. For $M=32$, the linear extrapolation (red curve with open 
symbols) is consistent, as expected, with the corresponding result for $M=1$
(red curve with filled circles). However a better $\chi^2$ would be obtained
with a quadratic extrapolation, which in turn would require a sudden upturn
at very small $\tau$ to recover the correct zero time-step limit.
}
\label{fig:sizeconsistency_e}
\end{figure}

From the results reported in Fig.~\ref{fig:sizeconsistency_e}, we observe that 
the FN energies obtained with the algorithm of Ref.~\cite{C06} significantly 
increase with $M$ and approach the energies obtained with the LA algorithm.
Already with a system with 32 oxygen atoms, the FN energies at 
$\tau=0.1$ obtained with these two approaches become equivalent within 
the error bars. The energies given by the two algorithms must however 
extrapolate to different values as the time-step goes to zero\cite{C06}.
The lower panel of Fig.~\ref{fig:sizeconsistency_e} shows that for $M=32$, 
in particular, they have to depart from each other within a tiny time-step 
interval near the origin.
Because of this behavior, the algorithm of Ref.~\cite{C06} is bound to have a 
problematic extrapolation to the zero time-step limit for large enough systems.

\subsection{Size-consistent formulations: version 1}

To address this problem, the original algorithm of Ref.~\cite{C06} can be easily reformulated in a size-consistent manner by observing that the
transition $T^{\rm NL}$ (Eq.~\ref{T^NL}) can be factorized as
\begin{eqnarray}
T^{\rm NL}({\bf R}',{\bf R},\tau)&=&\frac{\Psi_{\rm T}({\bf R}')}{\Psi_{\rm T}({\bf R})}\prod_{i=1}^N\langle {\bf r}_i'|e^{-\tau{\nu}^{\rm NL}}|{\bf r}_i\rangle\nonumber\\
                                 &=&\prod_{i=1}^N\frac{\Psi_{\rm T}({\bf r}'_1\ldots{\bf r}'_i,{\bf r}_{i+1}\ldots)}{\Psi_{\rm T}({\bf r}'_1\ldots{\bf r}'_{i-1},{\bf r}_i\ldots)}\langle {\bf r}_i'|e^{-\tau{\nu}^{\rm NL}}|{\bf r}_i\rangle\nonumber\\
                                 &=&\prod_{i=1}^N\left(\delta_{{\bf r}'_i{\bf r}_i}-\tau v_{{\bf r}'_i{\bf r}_i}\right)\nonumber\\
                                 &=&\prod_{i=1}^N t^{\rm NL}({\bf r}'_i,{\bf r}_i,\tau)\,,
\end{eqnarray}
where $\nu^{\rm NL}$ is the non-local potential acting only on one electron due to all the atomic centers so that
the total non-local potential is given by the sum over the electrons, 
$\langle {\bf R}'|{\cal V}^{\rm NL}|{\bf R}\rangle=\sum_i \langle {\bf r}_i'|{\nu}^{\rm NL}|{\bf r}_i\rangle$.
We defined the matrix element $v_{{\bf r}'_i{\bf r}_i}$ as
\begin{eqnarray}
v_{{\bf r}'_i{\bf r}_i}=\frac{\Psi_{\rm T}({\bf r}'_1\ldots{\bf r}'_i,{\bf r}_{i+1}\ldots)}{\Psi_{\rm T}({\bf r}'_1\ldots{\bf r}'_{i-1},{\bf r}_i\ldots)}\langle {\bf r}_i'|{\nu}^{\rm NL}|{\bf r}_i\rangle\,.
\end{eqnarray}
The transition $t^{\rm NL}({\bf r}'_i,{\bf r}_i,\tau)$ displaces the $i$-th electron over the grid of quadrature points 
generated by considering only the $i$-th electron and all pseudoatoms which host the $i$-th electron in their core region.

The FN approximation is applied separately to each single-electron 
transition as 
\begin{eqnarray}
t_{\rm FN}^{\rm NL}({\bf r}',{\bf r},\tau)=\delta_{{\bf r}'{\bf r}}-\tau v^-_{{\bf r}'{\bf r}}\,,
\end{eqnarray}
where $v^\pm_{{\bf r}'{\bf r}}$ is defined in analogy to the case of the total non-local potential so that
only the positive transition elements are kept in the transition matrix. 
In this formulation, the 
third step of the DMC algorithm detailed above consists of a loop over the electrons where each
electron is subsequently moved according to the single-electron transition, $t_{\rm FN}^{\rm NL}$.
Therefore, while in the original algorithm of Ref.~\cite{C06} the 
configuration generated in the $T_{\rm FN}^{\rm NL}$ step differs from the starting configuration 
only in the coordinate of one electron, the configuration resulting from this size-consistent move will 
generally change in more than one electronic coordinate and the number of electrons being moved will increase
with the size of the system.

To understand that the drift-diffusion-branching steps in the original algorithm do not need to be modified, we observe that 
the expression of the effective Hamiltonian ${\cal H}_{\rm eff}$ we are working with is the same as in
Eq.~\ref{H_eff} in the limit of $\tau$ going to zero. 
In particular, the sign-flip term obtained by summing all discarded terms $v^+_{{\bf r}'{\bf r}}$ 
over all the electrons is equal to the sign-flip term in ${\cal V}_{\rm eff}^{\rm loc}$ (Eq.~\ref{V_eff^loc})
to zero-order in $\tau$.  Similarly, we have that, to order $\tau$,
\begin{eqnarray}
\label{v1}
\prod_{i=1}^N\sum_{{\bf r}'_i}t_{\rm FN}^{\rm NL}({\bf r}'_i,{\bf r}_i,\tau)
\approx \sum_{{\bf R}'}T_{\rm FN}^{\rm NL}({\bf R}',{\bf R},\tau)\,,
\end{eqnarray}
and we recover the same branching factor as in the original algorithm. Therefore, both algorithms
extrapolate to the same limit at zero time-steps. We will refer to this improved algorithm as ``DMC version 1'', to
distinguish it from another size-consistent version we will define later in this section.
We stress here that by ``version 1'' we do not only mean the use of the product of 
single-particle $t_{\rm FN}^{\rm NL}$ in step 3, but also the symmetrization of the weights in step 2, 
as described in Eq.~\ref{sym_branching}, where the initial configuration is taken before the diffusion process (step 1),
and the final is the one after step 1.

The acceptance as a function of time-step using the size-consistent DMC 
algorithm (version 1) is shown in Fig.~\ref{fig:jwalkalize}. We only report 
the result obtained with $M=1$ as the curves for the other system sizes 
are exactly equivalent within statistical error. This finding can 
be easily understood since the probability of moving a given electron 
on the grid generated by considering all centers will be practically 
the same as the probability computed using only the atom close to the 
electron as all other centers are at least 30 $\AA$ far apart. Therefore, 
in the new size-consistent algorithm, when more atoms are added to increase
the size of the system, the loop over the electrons will ensure that each 
electron attempts a move around its closest center. The acceptance remains 
therefore constant as more oxygen atoms are added.  In a more realistic 
systems (e.g.\ with closer oxygen atoms), there will be a weak dependence 
on the size of the system but, after enough atoms have been added for most 
electrons to experience an equivalent environment, the acceptance will 
become independent on the system size given the short-range nature of 
the non-local components of the pseudopotentials.

The FN energies obtained for the oxygen systems with this size-consistent 
algorithm are compared in Fig.~\ref{fig:sizeconsistency_e} with the 
results of the LA and of the original algorithm.
We observe that, as expected, the FN energies of the size-consistent 
algorithm extrapolate to the same value as the original algorithm as 
$\tau$ goes to zero. On the other hand, while the size-consistent FN energies 
are close to the values obtained with the original method for the smallest, 
one-atom system, the FN results obtained by the two methods depart from 
each other as the system size increases. Importantly, the FN energies of 
the size-consistent scheme do not approach the LA results for large systems 
at finite $\tau$ and their extrapolation to zero time-step is therefore as 
smooth for large as for small systems.

\subsection{Size-consistent formulations: version 2}

An alternative scheme to address the size-consistent problem of the original algorithm of Ref.~\cite{C06}
can be obtained through a different route by starting from Eq.~\ref{T^NL}, 
and breaking it up in $N$ terms with time-step of $\tau/N$, such that:
\begin{equation}
\label{v2}
\sum_{{\bf R}'} T_{\rm FN}^{\rm NL}({\bf R}',{\bf R},\tau) = \sum_{{\bf R}_1 \cdots {\bf R}_{N}}  
\prod_{i=1}^N T_{\rm FN}^{\rm NL}({\bf R}_i,{\bf R}_{i-1},\tau/N)\,,
\end{equation}
with ${\bf R}_N={\bf R}'$, ${\bf R}_0={\bf R}$, and the sum over the quadrature points sampled by the 
chain $\{ {\bf R}_0,\ldots,{\bf R}_i,\ldots,{\bf R}_N \}$
generated during the random walk. This is another way to evaluate the quantity in Eq.~\ref{v1}. 
The difference is that 
Eq.~\ref{v2} involves a product of $N$ all-electron factors, while Eq.~\ref{v1} is a
factorization of $N$ single-electron terms. Both will avoid the saturation of the 
acceptance probability of the non-local Green's function $T_{\rm FN}^{\rm NL}$, 
and therefore they will ensure a size-consistent time-step error.
Since Eq.~\ref{v2} requires the calculation of \emph{all} matrix elements 
$V_{{\bf R}',{\bf R}}$ each time, it is more convenient to split the $N$ factors in such a way that 
the diffusion move involving the $i$-th electron could be placed between the $i-1$-th and $i$-th factor,
and the corresponding branching weight
updated as a product of subsequent single-electron components:
\begin{eqnarray}
\prod_{i=1}^N\,\exp\left[-\frac{\tau}{N} E_{\rm L}({\bf r}'_1\ldots{\bf r}'_i,{\bf r}_{i+1}\ldots,{\bf r}_N)\right]\,,
\end{eqnarray}
where we can exploit the knowledge of $V_{{\bf R}',{\bf R}}$ to compute also $E_{\rm L}$ for every single-electron move.
We will call this algorithm ``DMC version 2''~\cite{rqmc}. It consists of the following steps:
\begin{itemize}
\itemsep0ex
\item[1.] Diffusion move of the $i-$th electron.
\item[2.] The weight of the walker is multiplied by the branching factor 
$\exp\left[-\frac{\tau}{N} E_{\rm L}({\bf r}'_1\ldots{\bf r}'_i,{\bf r}_{i+1}\ldots,{\bf r}_N)\right]$.
\item[3.] The walker moves to ${\bf R}''$ according to the transition 
probability $T^{\rm NL}_{\rm FN}({\bf R}'',{\bf R}',\tau/N)/\sum_{{\bf R}'''}T_{\rm FN}^{\rm NL}({\bf R}''',{\bf R}',\tau/N)$
which involves \emph{all} the electrons.
\end{itemize}
In contrast to the original algorithm, and the ``DMC version 1'', 
these three steps need to be performed inside a loop over the electrons. 
In the ``version 2'' formulation of the DMC size-consistent algorithm, 
each electron drifts and diffuses in a time $\tau$ and the branching factor is 
updated at each SE move with the total local energy $E_{\rm L}$ and 
time $\tau/N$ in the exponent. After each SE branching update,
a non-local transition is performed with 
$T^{\rm NL}_{\rm FN}({\bf R}',{\bf R},\tau/N)$, 
where one electron among {\it all} electrons is displaced over the grid 
of quadrature points obtained by considering all electrons and all pseudoatoms.
Therefore, the electron displaced in the non-local move may differ from 
the electron which is currently being moved in the drift-diffusion step. 

\section{LRDMC and non-local pseudopotentials}
\label{Sec:lrdmc}

The main difference between the effective DMC Hamiltonian reported in Eq.~\ref{H_eff}
and the LRDMC one is the kinetic operator ${\cal K}$. In the LRDMC approach, ${\cal K}$ is 
replaced by a discretized Laplacian and treated on the same footing as ${\cal V}^{\rm NL}$.
In the original formulation,
the discretized Laplacian is a linear combination of two discrete operators 
with incommensurate lattice spaces $a$ and $a'$,
introduced to sample densely the continuous space by performing discrete moves 
whose length is either $a$ or $a'$.
This method can be simplified by noticing that all the continuous 
space can be visited  using  only  a single  displacement length $a$, 
provided we randomize the direction of the Cartesian coordinates each time 
the electron positions are updated.
The randomization of the lattice mesh is similar to the well established approach used 
to perform the angular integration in the non-local part of the pseudopotential~\cite{fahy}.
Therefore, in the LRDMC approach, we can extend the definition of
the kinetic part by including both the discretized Laplacian and the
non-local part of the pseudopotentials.  The total non-local operator reads:
\begin{equation}
\label{k_operator}
{\cal K}^a=
-\sum\limits_{i=1}^N \Delta^a_i(\theta_i,\phi_i)/2+ {\cal V}^{NL}  \,,
\end{equation}
where $\Delta^a_i(\theta_i,\phi_i)$ is the Laplacian acting on the 
$i$-th electron and discretized to second order so that $\Delta^a_i (\theta_i,\phi_i) = \Delta_i + O(a^2)$.
The discretized Laplacian is computed in a frame rotated by the angles
$\theta_i$ and $\phi_i$, which are chosen randomly and independently of 
the ones used to compute ${\cal V}^{NL}$.
In this formulation, we need to evaluate only 6 off-diagonal elements of the Green's function instead of 
12 as in the original algorithm, gaining a speed-up of a factor of 2 in full-core calculations and of 
${(12+\mathrm{n_\mathrm{quad}})}/{(6+\mathrm{n_\mathrm{quad}})}$ with pseudopotentials, where 
$\mathrm{n_\mathrm{quad}}$ is the number of quadrature points per electron~\cite{heavy}.
We notice that, with this simplification, the LRDMC error 
in the extrapolation to the continuous limit depends on a single   
parameter $a$, and the method can therefore be compared fairly with the DMC 
approach where the discretization of the diffusion process also
depends on a single scale, i.e. the 
time-step $\tau$.

In the LRDMC choice of the Hamiltonian, we further regularize the
single-particle operator ${\nu}$, 
defined as the electron-ion Coulomb interaction in full-core atoms or the local part ${\nu}^{\rm loc}$ of the pseudopotential,
so that 
${\nu}({\bf r}_i) \to {\cal V}_i^a ({\bf R})$ as
\begin{equation} 
{\cal V}_i^a({\bf R}) = {\nu} ({\bf r}_i) - {  \frac{( \Delta_i  -  \Delta_i^a) 
\Psi_T({\bf R})}{2 \Psi_T({\bf R})} }\,,
\label{defvc}
\end{equation}
when acting on the $i$-th electron. The single-particle operator ${\nu}$ 
acquires therefore a many-body term and ${\cal V}_i^a({\bf R})$
depends on the all-electron configuration ${\bf R}$. The total potential term is then given by 
\begin{equation}
{\cal V}^a = \sum\limits_{i=1}^N{\cal V}_i^a + {\cal V}_{ee} + {\cal V}_{nn}\,,
\label{pot_tot}
\end{equation}
where no regularization is employed in the electron-electron ${\cal V}_{ee}$ 
and ion-ion ${\cal V}_{nn}$ Coulomb terms. 
This lattice regularization leads to an approximate 
Hamiltonian ${\cal H}^a = {\cal K}^a + {\cal V}^a$ 
which converges to the exact Hamiltonian 
as ${\cal H}^a = {\cal H}+a^2 \Delta {\cal H}$
for $a\to 0$, where we denote with $a^2 \Delta {\cal H}$ the 
$O(a^2)$ LRDMC error on ${\cal H}$. 

The lattice Green's function Monte Carlo algorithm can then be employed
to sample exactly the lattice regularized Green's function, $\Lambda - {\cal H}^a$, and project
the trial wave function $\Psi_T$ to the approximate ground state $\Psi_a^\mathrm{LRDMC}$
which fulfills the fixed-node constraint based on $\Psi_T$,
in complete analogy to the DMC framework~\cite{CFS05}.
Note that, since the
spectrum of ${\cal H}^a$ is not bounded from above, we need to take
the limit $\Lambda \rightarrow \infty$, 
which can be handled with no loss of
efficiency as described in Ref.~\onlinecite{lambda}.
The usual DMC Trotter breakup results in a time-step error,
while the LRDMC formulation yields a lattice-space error, but both approaches
share the same upper bound property and converge to the same projected FN energy
in the limit of zero time-step and 
lattice-space, respectively.

Since the discretized Laplacian and the non-local potential are treated on the same footing, and the
sampling of the Green's function is based on a sequence of single-particle moves generated both
from the Laplacian and the non-local part, the LRDMC is intrinsically size-consistent (in the sense
previously discussed for the DMC algorithm), and no modification is necessary to make the lattice-space bias independent
of the system size. It will depend however on the quality of the trial wave function in the way detailed below.

\subsection{Small $a^2$ correction for good trial function }

The regularization of the potential (Eq.~\ref{defvc})
in the definition of the lattice Hamiltonian ${\cal H}^a$ 
implies that the correction $\Delta {\cal H}$ satisfies:
\begin{equation} \label{p1}
 \Delta {\cal H} | \Psi_T \rangle =0.
\end{equation}
Using this property, we can estimate the leading-order error of the lattice regularization
by simple perturbation theory as
\begin{eqnarray}
E^a&=& E^0 + a^2 \langle   \Psi_0^\mathrm{LRDMC} |  \Delta {\cal H} | \Psi_0^\mathrm{LRDMC} \rangle
\nonumber \\
 &=&  E^0 + a^2 \langle   \Psi_0^\mathrm{LRDMC} - \Psi_T |  \Delta {\cal H} | \Psi_0^\mathrm{LRDMC} 
 -\Psi_T  \rangle 
\nonumber \\
   &=&  E^0 + O( a^2 | \Psi_0^\mathrm{LRDMC} - \Psi_T |^2 )  \label{smalla}\,,
\end{eqnarray} 
where $E^a$ is the expectation value of the Hamiltonian ${\cal H}^a$ on the 
approximate FN ground state $\Psi_a^\mathrm{LRDMC} $ and $E^0$ the estrapolated
value as $a\rightarrow 0$.
Thus, the approach to 
the continuous limit is particularly fast for good trial functions, 
namely for $\Psi_T$ close to the ground state solution, since
$\Psi_0^\mathrm{LRDMC}$ is a state with lower energy than $\Psi_T$ and
has to approach the ground state at least as $\Psi_T$ does.
The leading corrections to the continuous limit 
are  quadratic in the wave function error. This property is not easily generalized 
to  the usual DMC method  and, to our knowledge, has not been 
established so far.
 
\subsection{Well defined lattice regularization}

As in any lattice model, the Hamiltonian ${\cal H}^a$
has a finite ground state energy 
only if the potential ${\cal V}^a$ is always limited from below. 
If ${\cal V}^a( {\bf R}_0) = -\infty $ for some configuration ${\bf R}_0$, the 
variational state $\Psi({\bf R}) = \delta_{{\bf R},{\bf R}_0}$ will have unbounded negative 
energy expectation value and the ground state 
energy of ${\cal H}^a$ is not defined.
Unfortunately, the regularized potential ${\cal V}^a({\bf R})$ in Eq.~(\ref{defvc}) is not bounded from below when ${\bf R}$ belongs to 
the $(3N-1)$-dimensional nodal surface ${\cal N}$ defined by the equation $ \Psi_T({\bf R})=0$. 
To cure these divergences, we need to be able to establish when a configuration is close
to the nodal surface.
In the lattice regularized formulation, we can assign an electron position  
${\bf r}_i$ to the nodal surface, i.e.\ $ {\bf r}_i \in  {\cal N }_a$,
if $  \Psi_T ( {\bf r}_i + a \vec \mu)$ has  the opposite  sign of 
$ \Psi_T ({\bf r}_i)$ for at least one of the six points used to evaluate the finite difference Laplacian 
(i.e. $\vec \mu$ is one of the six unit vectors $\pm \hat x, \pm \hat y$ 
or $\pm \hat z$ of the reference frame randomly oriented according to the 
angles $\theta_i$ and $\phi_i$).
${\cal N}_a$ correctly defines the nodal surface ${\cal N}$ in the limit $a\to 0$.

With this definition  of nodal surface, we can
modify ${\cal V}_i^a$ so that it remains 
finite when $ {\bf r}_i \in  {\cal N}_a$: 
\begin{equation} \label{vpota}
\tilde {\cal V}_i^a ( {\bf R} ) = \left\{
\begin{array}{cc} 
 \mathrm{Max} \left[ {\nu} ( {\bf r}_i) , { \cal V}_i^a ( {\bf R}) \right] & 
 {\rm if } ~    {\bf r}_i \in { \cal N }_a   \\
 {\cal V}_i^a( {\bf R})   & {\rm otherwise } 
\end{array}\,.
\right.
\end{equation}
If $ {\bf r}_i  \notin {\cal N}_a$,
we use the original LRDMC definition of ${\cal V}_i^a$ since 
${\cal V}_i^a$ remains finite even when an electron approaches a nucleus
for trial functions which satisfy the electron-ion 
cusp conditions. If $ {\bf r}_i  \in {\cal N}_a$, we need to distinguish two cases. If the electron
is not close to a nucleus, the regularized ${ \cal V}_i^a$ can diverge negatively
 while ${\nu}({\bf r}_i)$ remains finite and, according to Eq.~\ref{vpota}, 
the potential $\tilde {\cal V}_i^a$  coincides with $\nu$.
If the electron is close to a nucleus in a full-core calculation,
both ${ \cal V}_i^a$ and ${\nu}({\bf r}_i)$ diverge, so we need to further 
regularize ${\nu} ({\bf r}_i)$ in the right hand side of Eq.~(\ref{vpota}) 
and use an expression bounded from below. In this particular case, we choose to
replace the divergent electron-ion contribution  $ -Z/ | {\bf r}_{in} |$
in ${\nu} ({\bf r}_i)$
with $ - Z/a $ whenever the electron-ion distance $ |{\bf r}_{in}| < a $.

If we employ the regularized potential $\tilde {\cal V}^a$ in the Hamiltonian ${\cal H}^a$, we no longer satisfy Eq.~({\ref{p1})
and, in principle, it is not possible to compute $E^a$ by averaging the local energy ${\cal H}\Psi_{\rm T}/\Psi_{\rm T}$.
However, the use of $\tilde {\cal V}^a$ introduces only 
negligible errors in the computation of $E^a$ because the regularization is adopted only in
a region of volume $S \times a$,
where $S$ is the area of the nodal surface ${\cal N }$. Since
both the trial and the LRDMC wave function vanish $\simeq a$ 
close to the nodal surface,
the finite lattice error  corresponds to averaging
$ ({\cal H}^a-{\cal H}) \Psi_T/\Psi_T (\propto a$)
over $\Psi_T \Psi_{FN} (\propto a^2$)
in a nodal region of extension $\propto a$.
If we collect these contributions, 
we find that the present regularization
introduces a bias in the nodal region which vanishes as $ a^4$ 
for $a \to 0$ and is always negligible compared to the dominant 
contribution $ O( a^2 | \Psi_0 - \Psi_T |^2 )$.
Moreover, since the regularization in Eq.~(\ref{vpota}) acts independently 
on each electron, 
it does not affect the size-consistent character of the algorithm, and
the energy of $N$ independent atoms at large distances is equal to 
$N$ times the energy of a single atom.
Therefore, we did not perform any LRDMC calculations for the oxygen systems since
the energy per atom as a function of $a$ is exactly independent of $N$.

\section{Performance of the proposed methods}
\label{Sec:performance}

An important point to address is the efficiency of our revised techniques. 
This involves not only the computational cost per Monte Carlo step, but
also the elimination of the discretization error (in time or space, as
appropriate) by extrapolation to the continuum limit. Indeed, a smaller 
and smoother bias enhances the overall efficiency, as does the knowledge
of the leading term in the discretization parameter.

\subsection{Time-step error}
We study the time-step error on the FN energy computed with the various algorithms discussed above,
using the Oxirane molecule (C$_2$H$_4$O) as a test case.  
Our aim here is in particular to assess the reduction of the time-step error
with respect to the original algorithm~\cite{C06}. 
In the DMC ``version 1'' this reduction is due to the symmetrization of the weights,
while in the DMC ``version 2'' it is due to the 
update of the branching factor after single-electron moves.
We employ non-local energy-consistent Hartree-Fock 
pseudopotentials~\cite{BFD07} for the oxygen and the carbon atoms in combination with the corresponding cc-pVDZ basis 
sets, and construct two single-determinant Jastrow-Slater wave functions of different quality. The first wave function
is built from B3LYP orbitals and a very simple electron-electron Jastrow factor of the form $b [1-\exp(-\kappa r_{ij})]/\kappa$,
where $b=1/2$ or $1/4$ for antiparallel- and parallel-spin electrons, respectively. The parameter $\kappa$ is optimized in 
energy minimization and is equal to 1.91.
The second wave function is characterized by a more sophisticated Jastrow factor comprising of 
electron-electron, and electron-nucleus terms, and all orbital and Jastrow parameters in the
wave function are optimized in energy minimization.

The top panel of Fig.~\ref{fig:Oxirane} shows results
obtained with the simple wave function. Consistently with previous 
studies on the water molecule~\cite{needs-water}, the LA energies 
extrapolate to a lower value (not necessarily variational)
than the original algorithm of Ref.~\cite{C06}, with a
smaller time-step error; symmetrization of the branching factor 
in the original algorithm is already sufficient to reduce 
the time-step error down to a value similar to that found in the LA.
As expected, given the small size of the system considered, the 
original and the size-consistent algorithm give nearly identical 
results, as shown here for its version 1 with AE branching.

The main result shown in the top panel of Fig.~\ref{fig:Oxirane} is the remarkable reduction 
of the time-step error obtained with a SE branching factor. The data shown 
in the Figure refer to version 2 of the size-consistent algorithm.
We also mention, without reporting the data, that 
when the branching factor is updated after SE moves, the symmetrization of
the local energy in the exponent does not improve the time-step error 
significantly.

\begin{figure}[t]
\includegraphics[width=1.0\columnwidth]{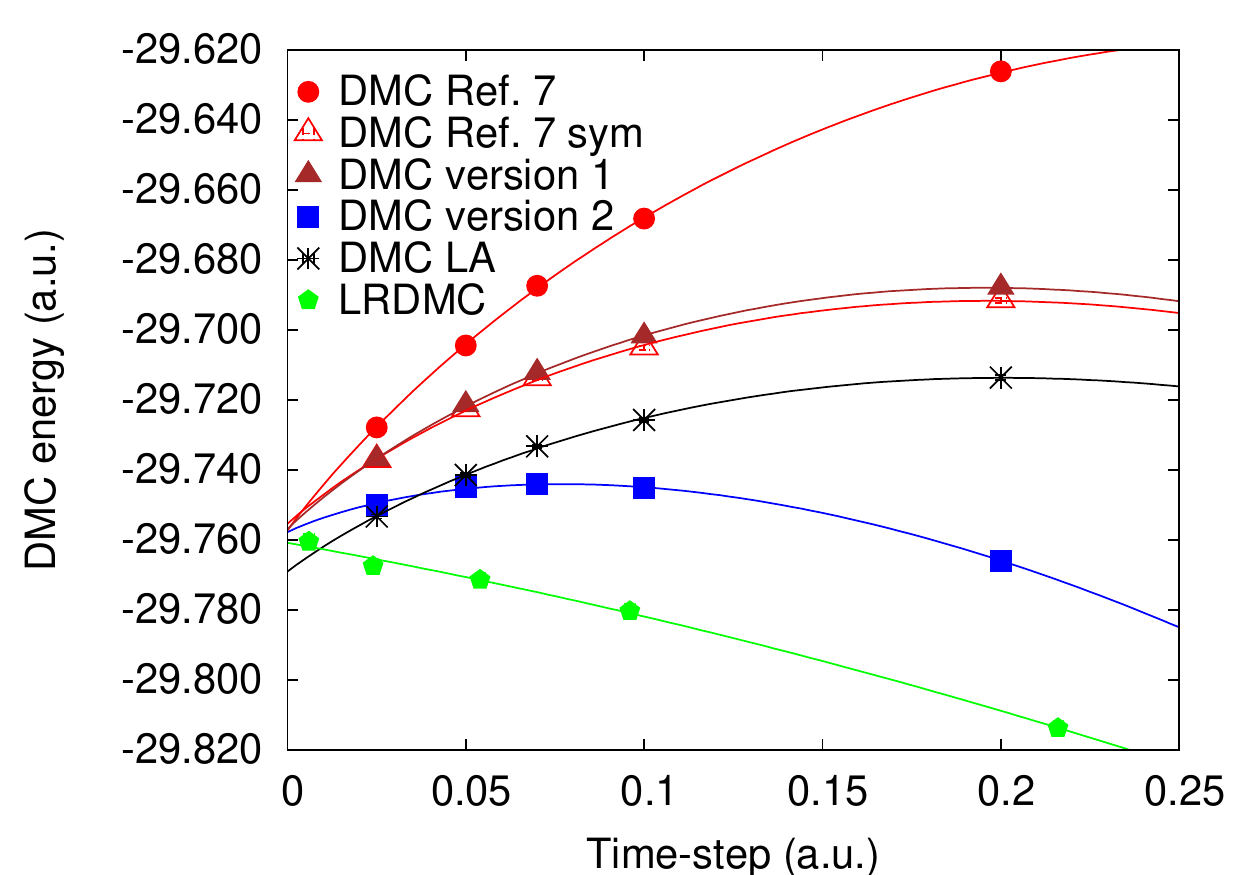}\\
\includegraphics[width=1.0\columnwidth]{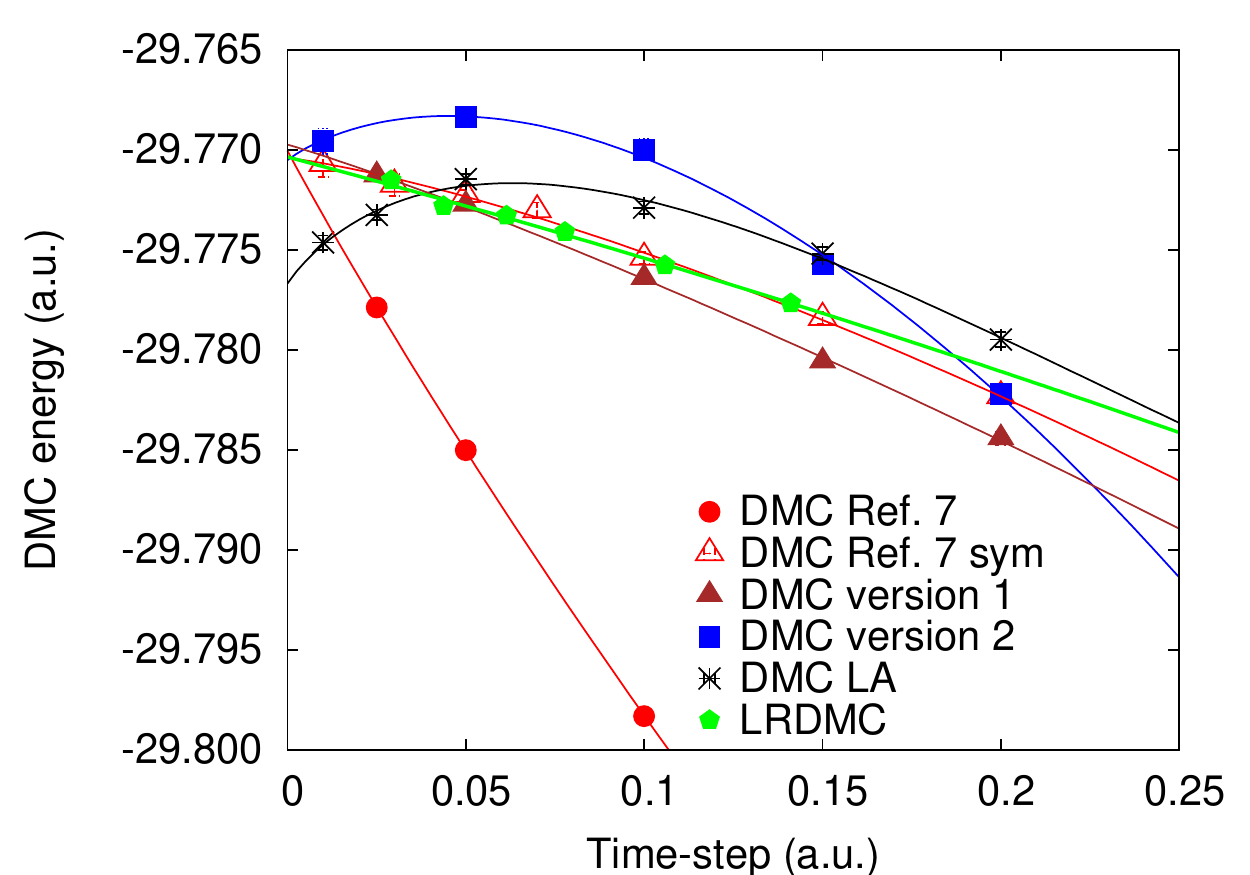}
\caption[]{FN energies as a function of time-step for the Oxirane (C$_2$H$_4$O) molecule, obtained using a simple (top) and a
more sophisticated (bottom) trial wave function. We employ different schemes, i.e. the original algorithm 
as in Ref.~\cite{C06} and with an improved symmetrized branching factor (``sym''), the two size-consistent 
approaches we proposed (``DMC version 1'', and ``DMC version 2''), the LA approach, and the LRDMC method.
The lattice-space has been mapped into the time-step via the relation $\tau = 0.6 ~ a^2$, which
guarantees the same autocorrelation time between the ``DMC version 1'' and the ``LRDMC'' method
in this particular case.}
\label{fig:Oxirane}
\end{figure}

The improvement obtained with a SE branching factor, however, is strongly 
dependent on the quality of the trial function. The lower panel of 
Fig.~\ref{fig:Oxirane} shows results obtained with the more sophisticated 
wave function. We still find a lower energy with the LA (with its possibly 
problematic behavior at very small time-step), and a large time-step error 
with the original algorithm of Ref.~\cite{C06}. All the other cases, however,
display similar behavior, or at least comparable quality, in terms of the 
time-step error.

Also the LRDMC energy values are reported in Fig.~\ref{fig:Oxirane}, where the lattice-space
has been converted into time-step based on the equal auto-correlation time between
Monte Carlo generations in the DMC and LRDMC algorithms. This is the fairest mapping
since it keeps the final statistical error equivalent for the same sample length.
In this case, it gives $\tau \sim 0.6 ~ a^2$.
One can see that the LRDMC 
energies are always converging from below in a monotonic way, usually 
easier to extrapolate than the corresponding DMC energies.

In order to make a more quantitative analysis of the predictions reported 
in Section~\ref{Sec:lrdmc} for the lattice-space error, we studied 
the lattice-space extrapolation of the Oxirane molecule with the DFT-B3LYP Slater determinant, 
and Jastrow factors going from the simple
2-body one, to the most complicated comprising of one-, two-, and three-body terms.
The results are reported in Fig.~\ref{fig:lrdmc}. For good trial wave functions,
a reliable extrapolation can be obtained even by using very large values of 
$a$, where small statistical errors can be obtained with 
much less computational effort. 
Also, the FN energies are basically independent of the shape 
of the trial wave function already for a rather simple Jastrow with 1-body and 2-body terms, 
implying that the ``locality error'' becomes negligible
in the variational formulation even for not-so-accurate trial wave functions. This consideration applies also
to the DMC variational energies, since the zero-lattice-space zero-time-step limits are equivalent.
\begin{figure}[t]
\includegraphics[width=1.0\columnwidth]{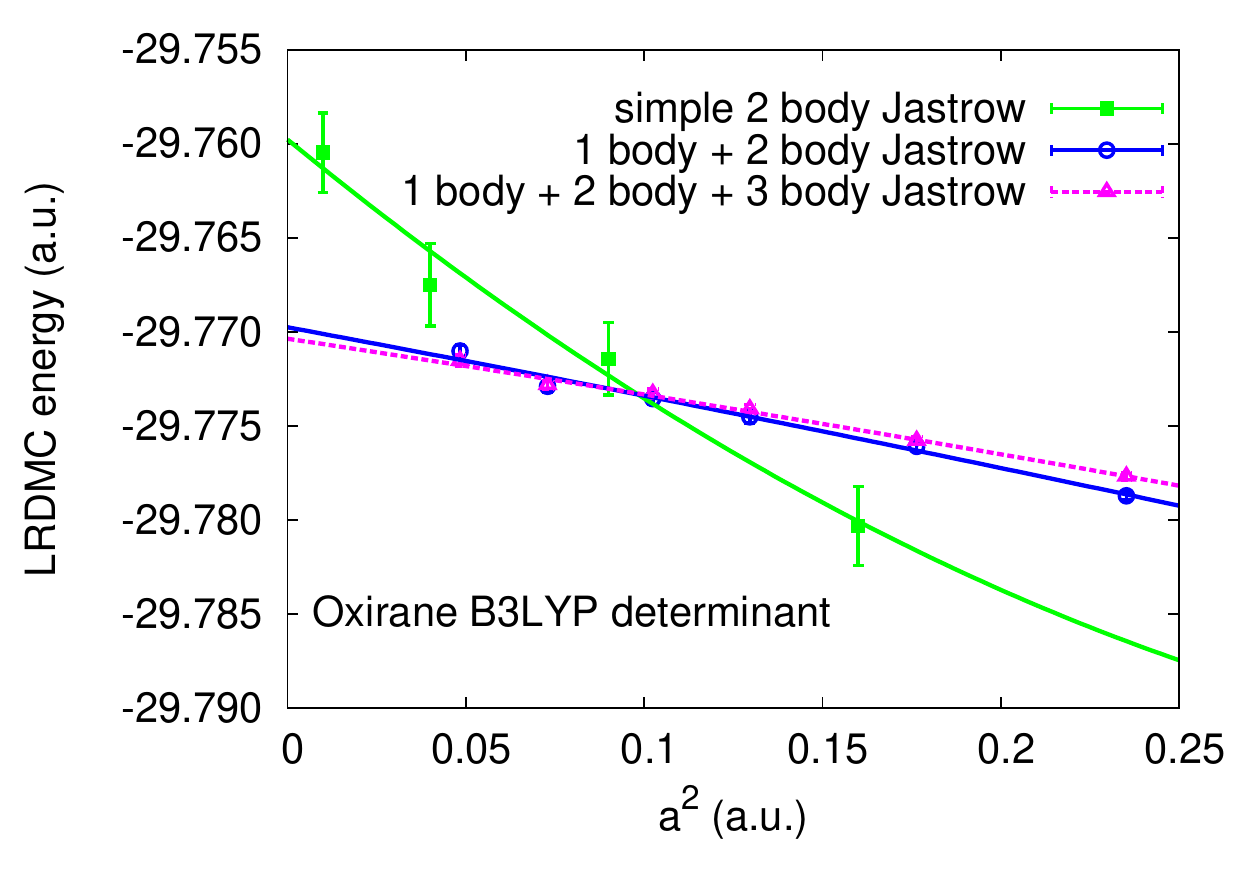}
\caption[]{FN LRDMC energies as a function of lattice-space $a$ for the Oxirane (C$_2$H$_4$O) molecule, obtained using 
three types of Jastrow factors. The fitting curves include a quadratic and quartic term, namely 
$f(a) = E_0 + b a^2 + c a^4$. 
The finite lattice space error improves dramatically with a better wavefunction.
For the simple 2-body Jastrow, $b \approx -0.15$, while for the most accurate Jastrow factor 
$b' \approx -0.03$. 
The ratio of the variances of the two trial wave functions is roughly equal to $b/b'$, 
in agreement with Eq.~(\ref{smalla}). 
}
\label{fig:lrdmc}
\end{figure}

\subsection{Relative efficiency}

In all the methods presented here, there is an extra computational cost 
per Monte Carlo step with respect to
the standard DMC with LA since an extra step is needed in order to sample
correctly the Green's function related to the non-local pseudopotentials.
However, we have seen that, in \emph{all} the variational methods,
the non-local pseudopotential operator
will displace only one electron a time since the non-local pseudopotential gives a
one-body contribution to the Hamiltonian. This means that, in order to
update all the quantities needed by the simulation as the wave function ratios,
$V_{{\bf R}',{\bf R}}$, the gradients and Laplacian terms, one can exploit 
the Sherman-Morrison algebra, which scales as $N^2$. For instance, to
update the non-local term $V_{{\bf R}',{\bf R}}$, as well as the ${\cal K}^a$ in the LRDMC,
one employs the same algebra as the one used to update the gradient (i.e. the drift term) 
in the standard DMC with importance sampling. After a single particle move, 
the cost to \emph{fully} update $V_{{\bf R}',{\bf R}}$ scales as $\mathrm{n_\mathrm{off}}~ N^2$
where $\mathrm{n_\mathrm{off}}$ is the number of non-local mesh points per electron ($\mathrm{n_\mathrm{off}}=\mathrm{n_\mathrm{quad}}$
in DMC with non-local pseudopotentials and $\mathrm{n_\mathrm{off}}=\mathrm{n_\mathrm{quad}}+6$ in LRDMC with
non-local pseudopotentials and a single lattice-space in the Laplacian). 
In the size-consistent DMC (both ``version 1'' and ``version 2''), the pseudopotential move
has to be performed $N$ times in a single time-step $\tau$, so the overall cost per time-step coming
from the pseudopotential operator is $\eta ~ \mathrm{n_\mathrm{off}} ~ N^3$, where $\eta$ is the acceptance ratio of
the non-local part. Since $\mathrm{n_\mathrm{off}} \approx 20$, 
and $\eta \approx 0.1$ at convergence (see Fig.~\ref{fig:jwalkalize}),
it is clear that the DMC ``version 1'' will be only a prefactor $\approx 2$ slower than the
standard DMC with LA. The ``version 2'' might be slightly slower than the ``version 1'' 
since it requires the calculation
of the local energy after each single-particle move, but again the difference will be just a prefactor.
The LRDMC approach is the slowest because the total number of operations 
in a cycle with $N$ single-electron updates of the local energy 
takes $ (10+\mathrm{n_\mathrm{quad}}) /( 4 +\mathrm{n_\mathrm{quad}})\le 2.5$ more operations 
(the worst case is for full-core calculations when $\mathrm{n_\mathrm{quad}}=0$).
Moreover, there is also an additional slowing down compared to the  
DMC ``version 1'' approach because, in the latter case,
all operations involving the local energy can be done at the end of a cycle 
and cast in a very efficient form using matrix-matrix multiplications 
of size $\sim N$. These operations, for large $N >\simeq 1000$, 
 can be much more efficient than 
single-electron matrix updates (by a factor ranging from 2 to 20, depending 
on the computer hardware and software).  At present, it is 
difficult to estimate how much slower LRDMC will be on a particular machine~\cite{efficiency_LRDMC},
also considering that further algorithmic and software developments are expected
in the near future, which should allow faster updates. 
However,  even though LRDMC is certainly slower, it has the advantage of a much smoother lattice-space extrapolation as discussed above.

\section{Conclusions}
In conclusion, we have introduced important developments in the 
DMC and LRDMC methods in the context of electronic structure simulations 
with non-local pseudopotentials. 

We have explained how to modify the
DMC variational formulation for non-local potentials of 
Ref.~\cite{C06} in order to make it size-consistent. 
We have shown that, for large systems, the original algorithm~\cite{C06}
will depart from the usual localization approximation
only for small time-steps, making the zero time-step extrapolation
possibly problematic. Instead, the two DMC algorithms presented here, based 
on a more accurate Trotter break-up for the non-local operator and 
a better branching factor, have a smaller and size-consistent time-step error.
The DMC version 1, which features a single-particle representation 
of the non-local operator and a branching factor 
symmetric with respect to the application of the diffusion operator, 
is straightforward to implement in the existing codes.
The DMC version 2 is closer to the LRDMC spirit, since
the non-local part is further split in $\tau/N$ factor always 
acting on the all-electron configuration, and the branching factor
is accumulated after every single-particle move.
The latter version can give an even better time-step error 
(order $O(\tau/N)$ in the non-local part), 
particularly for relatively poor wave functions.
In general, it is slightly more time consuming than 
the version 1, since it requires the evaluation of the full non local matrix 
after every single-particle move. 

We have made significant progress also in the LRDMC approach. 
In the present formulation, it is no longer
necessary to use two lattice meshes to randomize the electron position, 
but a single lattice space $a$ is sufficient, provided the orientation of the 
Cartesian coordinates of the discretized Laplacian is changed randomly 
during the diffusion process.
We have defined a better lattice regularization of the Hamiltonian in order
to have always a potential bounded from below, with a cutoff depending on $a$.
This leads to a well defined and size-consistent lattice-space extrapolation since,  in the 
$a \rightarrow 0$ limit, we recover the variational expectation value of the 
continuous Hamiltonian with a lattice space error whose leading term is quadratic in $a$.
Moreover, we showed that the prefactor of the $a^2$ term vanishes quadratically in 
$|\Psi_0 - \Psi_T|$. 
Therefore, for good wave functions,
the extrapolation to the $a\to 0$ limit is particularly 
rapid and smooth with a computational effort $\propto 1/a^2$.
The DMC error appears instead to be less correlated to the 
quality of the guiding function and may display a turn-down behavior for small time-steps (observed
here and elsewhere~\cite{needs-water}), which makes the time-step extrapolation 
much harder than in the LRDMC lattice-space approach. 
Regarding the computational cost, the LRDMC approach is slower but the overall efficiency is comparable
to the two variational and size-consistent DMC formulations presented here since LRDMC 
allows one to work with large values of $a$ due to the robust extrapolation to the zero lattice-space limit.

\acknowledgments

CF acknowledges the support from the Stichting Nationale Computerfaciliteiten
(NCF-NWO) for the use of the SARA supercomputer facilities. MC thanks the
computational support provided by the NCSA of the University of Illinois at Urbana-Champaign. SS and SM acknowledge support from CINECA and COFIN'07.


\begin{thebibliography}{99}
\bibitem{heavy}For heavy elements treated either as full-core atoms
or with small-core pseudopotentails, the two-mesh discretization might 
remain a better choice since it allows a more efficient sampling of the 
different length scales in the core and valence regions.
\bibitem{efficiency_LRDMC} The relative efficiency between LRDMC and ``DMC version 1''  within the same machine
(Intel quad-core) and the same code~\cite{turbo} is less than a factor $2$ for the cases shown in this work.
\end{thebibliography}
\end{document}